\documentclass{ws-ijmpcs}

\begin{document}

\markboth{Christian M. Fromm}
{Simulation of Shock-Shock interaction in parsec-scale jets}

\catchline{}{}{}{}{}
%
%%%%%%%%%%%%%%%%%%%%%%%%%%%%%%%%%%%%%%%%%%%%%%%%%%%%%%%%%%%%%%%%%%%%

\title{Simulation of shock-shock interaction in parsec-scale jets}

\author{Christian M. Fromm$^{1}$}

\address{cfromm@mpifr.de}

\author{Manel Perucho$^{2}$, Eduardo Ros$^{2,1}$, Petar Mimica$^{2}$, Tuomas Savolainen$^{1}$, Andrei P. Lobanov$^{1}$\\ and J. Anton Zensus$^{1}$}

\address{$^{1}$Max Planck Institute for Radio Astronomy, Auf dem H\"ugel 69,53121 Bonn, Germany\\
$^{2}$Department d'Astronomia i Astrof\'isica, Universitat de Val\`encia, C/Dr. Moliner 50, Burjassot (Val\`encia) 46100, Spain}

\author{}

\maketitle

\begin{history}
\received{Day Month Year}
\revised{Day Month Year}
\end{history}

\begin{abstract}
The analysis of the radio light curves of the blazar CTA\,102 during its 2006 flare revealed a possible interaction between a standing shock wave and a traveling one. In order to better understand this highly non-linear process, we used a relativistic hydrodynamic code to simulate the high energy interaction and its related emission. The calculated synchrotron emission from these simulations showed an increase in turnover flux density, $S_{m}$, and turnover frequency, $\nu_{m}$, during the interaction and decrease to its initial values after the passage of the traveling shock wave.  

\end{abstract}

\ccode{PACS numbers: 11.25.Hf, 123.1K}

\section{Introduction}	

The blazar CTA\,102 ($z$=1.036) underwent a major flux density outburst in April 2006
reaching a historical value of 10\,$\mathrm{Jy}$ at 37\,$\mathrm{GHz}$\cite{fro11b}. The 
analysis of the single dish light curves revealed an unexpected double hump structure in
the turnover frequency - turnover flux density $\left(\nu_{m}-S_{m}\right)$ plane. A detailed
modeling of this flare with the shock-in-jet model\cite{mar85} suggested a possible traveling-standing
shock interaction as the origin of the flux density increase\cite{fro11}. This scenario is supported by
studies of the 15\,$\mathrm{GHz}$ and 43\,$\mathrm{GHz}$ parsec-scale morphological changes in the source
from very-long-baseline interferometry (VLBI) images. These 
studies show a stationary feature at a distance of 0.2\,$\mathrm{mas}$ (36\,$\mathrm{pc}$ de-projected)
away from the radio core (taken as the brightest feature at the base of the jet) during the period 2003 -2011. 
At the time of the 2006 flare, feature ejected from the jet base passed through
this stationary one, increased the observed flux densities and dragged the component outwards\cite{fro11b}.
Standing shocks or re-collimation shocks arise naturally in over-pressured jets and are characterized 
by a local increase in pressure, density and magnetic field\cite{dal88}. Relativistic hydrodynamical 
simulations of superluminal sources have shown that the interaction of the standing shock with a traveling
one could lead to the above mentioned observational effects\cite{gom97}. 

\section{Simulation}
For our simulation we used a relativistic hydrodynamic code\cite{per07}, based on a high-resolution shock-capturing scheme for 
relativistic hydrodynamics\cite{mar99}. We simulated a relativistic jet consisting of relativistic electrons and sub-relativistic 
protons using an ideal gas equation of state with a constant adiabatic exponent $\gamma=13/9$.
The initial jet radius was set to $R_{j}=0.3\,\mathrm{pc}$, with 
velocity $v_{j}=0.99652\,\mathrm{c}$, Mach number $M_{j}=3$, density $\rho_{j}=3.34\times 10^{-25}\,\mathrm{g/cm^{3}}$ and pressure 
$p_{j}=6.0\times 10^{-5}\,\mathrm{dyne/cm^{2}}$. The ambient medium is assumed to be constant along the jet with density 
$\rho_{a}=1.67\cdot10^{-24}\,\mathrm{g/cm^{3}}$ and pressure $p_{j}=2.0\cdot10^{-5}\,\mathrm{dyne/cm^{2}}$, which leads to an over-pressure 
of the jet of $d_{k}=P_j/P_a=3$. The numerical resolution was set to 16 cells per jet radii.This set of parameters corresponds to a jet with a kinetic power $L_j\sim 3.0\times 10^{45}\,\mathrm{erg/s}$, which is a typical value for FR II galaxies. 
After the jet reached a steady state, we injected a perturbation with pressure and density values four times larger than the values 
of the jet. For the duration of the injection event 
we assumed 3 months which was comparable to the observed duration of the increase of the flux density in CTA\,102 during the 2006 
radio flare. From the pressure, density, and fluid velocity distributions obtained, we calculated the synchrotron emission following 
refs.~\refcite{mim09,mim10} and references therein, assuming that the magnetic energy density is a fraction $\epsilon_{B}=0.1$ of the equipartition 
magnetic field and that a fraction $\epsilon_{e}=0.1$ of the shock energy is used to accelerate the non-thermal particles. The results of this simulation are
presented in Fig. \ref{fig1} and will be described in the following section.

\section{Results}
\begin{figure}[h!]
\centerline{\psfig{file=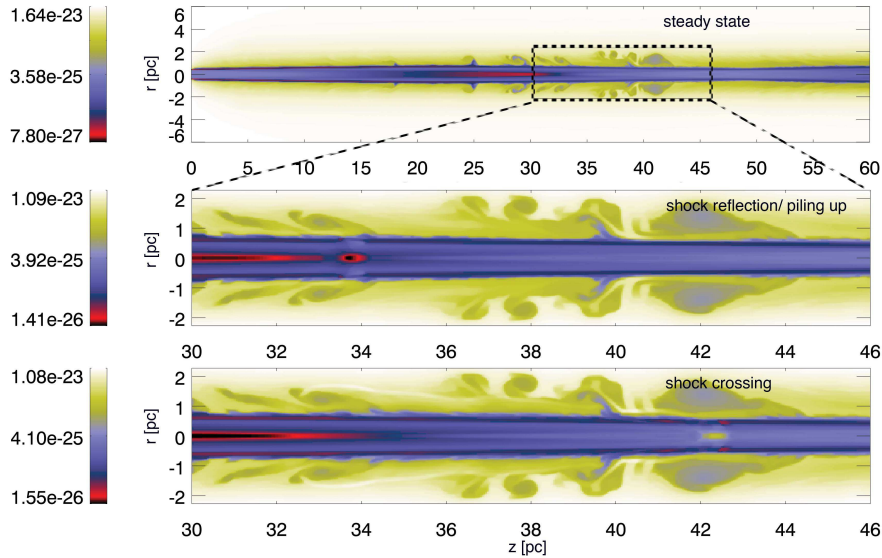,width=12cm}}
\vspace*{2pt}
\caption{Results of our RHD simulation of the interaction between a traveling and a stationary shock in a relativistic jet. Shown are the logarithm of the rest mass density for the steady-state jet (top panel), partial reflection of the traveling shock wave at the onset of the re-collimation shock (middle panel) and the crossing of re-collimation shock by the traveling shock (bottom panel) Notice the increased density value during the shock-shock interaction.}
\label{fig1}
\end{figure}

Due to the initial over-pressure in the jet, the fluid expanded radially after leaving the jet nozzle. During this process, a rarefied 
region forms close to the jet boundaries. In the inner jet regions, however, a conical shock from the injection generates a Mach disk 
around the jet axis at a distance of $z\sim10\,\mathrm{pc}$ from the injection. The final structure could be best described by an inner spine 
$\left(r<0.3\,\mathrm{pc}\right)$, characterized by density, pressure, and velocity values comparable to the ones at the jet nozzle, and 
an outer sheath $\left(0.3\,\mathrm{pc}< r <1\,\mathrm{pc}\right)$ with decreasing density and pressure profiles with distance and 
increasing fluid velocity, as expected from an adiabatic expansion. Downstream of the Mach disk, the fluid expands conically, showing a shallow 
acceleration and creating the first re-collimation shock at a distance of $z\sim33\,\mathrm{pc}$ from the jet nozzle. At this point, the fluid 
decelerates and the density and pressure values jump nearly to their initial values (factor 15 in pressure and factor 20 in density). The outer 
sheath shows a similar behaviour; the radial expansion reaches a maximum around a distance of $z\sim25\,\mathrm{pc}$ and starts to recollimate. 
This collimation gives rise to inward traveling sound waves which interact with the waves created by the collimation of the spine and form an 
extended region of increased pressure and density between $33\,\mathrm{pc}<z<50\,\mathrm{pc}$ (see top panel in Fig. \ref{fig1}). 
As mentioned before, we injected a perturbation after the jet reached a steady state.
In the following, we concentrate on the variations of pressure, density and velocity of the inner spine. The injected perturbation travels 
along a decreasing pressure and density profile producing an increase by a factor 4 in pressure and density (the jumps in 
pressure and density were decreasing with distance from the nozzle). However, at the position of the first re-collimation shock 
$\left(z\sim33\,\mathrm{pc}\right)$, there is a partial reflection of the perturbation and it can not pass through the standing shock at once (middle panel 
in Fig.~\ref{fig1}).
Instead, we found 
that the perturbation piles up to a certain pressure and density value before it can cross the re-collimation shock. The piled up material leads to 
an increase of a factor 12 in pressure and a factor of 6 in density at the location of the re-collimation shock, with respect to the steady-state
values (bottom panel in Fig.~\ref{fig1}).
At the same time, the standing shock is dragged downstream for a specific time before it is re-established at its initial position. 
We calculated the observed non-thermal emission using the known distance to the source, a viewing angle of $2.6^{\circ}$ and 
a spectral slope, $s=2.5$.  For the steady state jet we obtain a turnover frequency of $\nu_{m}\sim1\,\mathrm{GHz}$ and a turnover flux density 
of $S_{m}\sim 0.03\,\mathrm{Jy}$ at the position of the re-collimation shock and values of $\nu_{m}\sim 15 \,\mathrm{GHz}$ and 
$S_{m}\sim 0.24\,\mathrm{Jy}$ during the interaction with the traveling shock. After the passage of the traveling shock wave through the standing 
one, the emission values decreased back to the ones obtained for the steady state.

\section{Conclusions}
Our simulation of the interaction between a traveling shock and a stationary shock showed that this process can lead to a strong increase in the observed non-thermal 
radiation. However, the turnover values obtained during the shock-shock interaction 
are lower than the observed ones in CTA\,102 during the 2006 outburst ($\nu_{m}\sim50\,\mathrm{GHz}$, $S_{m}\sim8\,\mathrm{Jy}$).
This could be an indication for:% i) The density in relativistic jets is larger than 20\% that of the ambient medium (or the ambient medium is denser). ii) The influence of the magnetic field is higher than assumed i. e., $\epsilon_{b}>0.1$. iii) Shocks are more efficient in accelerating particles than assumed, i. e., $\epsilon_{e}>0.1$.
\begin{enumerate}
\item The density in relativistic jets is larger than 20\% that of the ambient medium (or the ambient medium is denser).
\item The influence of the magnetic field is higher than assumed i. e., $\epsilon_{b}>0.1$.
\item Shocks are more efficient in accelerating particles than assumed, i. e., $\epsilon_{e}>0.1$.
\end{enumerate}
To verify the assumptions above, two directions of research are planned. We will perform a deep analysis of the existing multi-frequency VLBI 
observations of CTA\,102 and an extended parameter study for the simulated shock-shock interactions.

\section*{Acknowledgments}
\small C.M.F. was supported for this research through a stipend from the International Max Planck Research School (IMPRS) for Astronomy and 
Astrophysics at the Universities of Bonn and Cologne. Part of this work was supported by the COST Action MP0905 Black Holes in a 
Violent Universe through STSM (MP0905-140211-004292). M.P. acknowledges financial support from the Spanish MICINN grants 
AYA2010-21322-C03-01, AYA2010-21097-C03-01 and CONSOLIDER2007-00050, and from the ``Generalitat Valenciana'' grant 
``PROMETEO-2009-103''. E. R. acknowledges partial support from MICINN grant AYA2009-13036-C02-02.

%\begin{thebibliography}{000} %for 3 digits
%\begin{thebibliography}{00}  %for 2 digits

\end{document}